\documentstyle[12pt,epsfig,citesort]{article}
\newcommand{\be}{\begin{equation}}
\newcommand{\ee}{\end{equation}}
\newcommand{\bdis}{\begin{displaymath}}
\newcommand{\edis}{\end{displaymath}}
\newcommand{\bx}{{\bf x}}
\newcommand{\bR}{{\bf R}}
\newcommand{\la}{\left\langle}
\newcommand{\ra}{\right\rangle}
\newcommand{\lp}{\left(}
\newcommand{\rp}{\right)}
\newcommand{\lab}{\left|}
\newcommand{\rab}{\right|}

\title{Statistics  of pressure and of pressure-velocity 
correlations in isotropic turbulence.}
\author{L. Biferale$^{1}$, P. Gualtieri$^2$ and F. Toschi$^{3}$}
\begin{document}
\maketitle
\centerline{$^1$Dipartimento di Fisica and INFM, Universit\`{a} 
di Tor Vergata,}
\centerline{ Via della Ricerca Scientifica 1, I-00133 Roma, Italy.}
\centerline{$^2$Dip. di Meccanica e Aereonautica, Universit\'a di Roma ``La Sapienza'',}
\centerline{Via Eudossiana 18, 00184, Roma, Italy.}
\centerline{$^3$University of Twente, Department of Applied Physics and }
\centerline{J.M. Burgerscentrum for Fluid Dynamics,}
\centerline{P.O. Box 217, 7500 AE, Enschede, The Netherlands}
\centerline{and INFM, Unit\'a di Tor Vergata,}
\centerline{Via della Ricerca Scientifica 1, I-00133 Roma, Italy.}
\medskip
   
\begin{abstract}
Some pressure and pressure-velocity correlations
in a direct numerical simulations of a three-dimensional
turbulent flow at moderate 
Reynolds numbers have been analyzed. 
We have identified a set
of pressure-velocity correlations which posseses  a good   scaling 
behaviour.  Such a class of  pressure-velocity correlations
are determined by looking at the energy-balance across any sub-volume
of the flow. \\
According to our analysis,
pressure scaling is determined by the dimensional
assumption that pressure behaves as a ``velocity squared'', unless 
finite-Reynolds effects are overwhelming. 
The SO(3) decompositions of  pressure
structure functions has also been applied 
in order to investigate anisotropic effects on the pressure scaling. 
\end{abstract}

\section{Introduction}
Scaling in turbulent flows is  one of the most challenging open issue in
fluid-dynamics \cite{frisch}.
Typical problems concern both the understanding of
the ideal case of isotropic and homogeneous turbulence in the limit of high 
Reynolds numbers \cite{iso1,iso2} or more realistic and applied 
situations with anisotropic and inhomogeneous statistics (for  recent 
examples see \cite{prl1,prl2,itamar1}).
In 1941,  Kolmogorov, used a clever application of dimensional analysis 
 to predict that the scaling of velocity increments in the inertial range
should have a power law
behaviour depending only on the averaged energy dissipation
 in the flow, $\epsilon$. Namely, for  structure function of order $p$, i.e. 
the p-th moment of a velocity difference across a distance $R$ we have:
\begin{equation}
S_q(R) = \la\lp v(x+R)-v(x)\rp^q\ra \sim  \epsilon^{q/3}\,R^{q/3}
\label{sf}
\end{equation}
with $\eta\ll r\ll L_0$ where $\eta$ is the dissipative scale and $L_0$ is
 the typical
external scale where forcing acts.  Let us notice that in (\ref{sf}) we have
 explicitly
neglected any tensorial structure in the velocity field such as to stress the 
typical
dimensional character of the Kolmogorov theory, i.e.
 the scaling properties
must be  the same 
for any observable which has the same physical dimension and which is built
 in terms
of local field increments, $\delta_r v(x) = 
v(x+r) -v(x)$.\\
Kolmogorov theory, as previously summarized, is quantitatively wrong. 
Experiments and numerical simulations  show a quantitative disagreement with 
the dimensional prediction  $p/3$ for the scaling exponents.
 For example,  the longitudinal velocity structure functions:
\begin{equation}
S^v_q\lp r\rp = \la \lab \lp v_i\lp\bx+\bR \rp-v_i\lp\bx\rp\rp\hat{R}_i\rab^q\ra 
\sim  R^{\zeta_v(p)}
\label{sf1}
\end{equation}
 show a power law behaviour 
with  a set of exponents $\zeta_v(p)$ non linear in $p$.  \\ 
The failure of the dimensional estimate $p/3$
goes under the name of anomalous scaling. \\
Many problems naturally arises as a consequence of the failure of the main 
Kolmogorov prediction. The main open problem  is  to find an analytical 
way to calculate from first principle the anomalous exponents, a
problem which is still out of control but for the  
 case of anomalous exponents characterizing the statistics of passive
 quantities advected by gaussian velocity fields 
\cite{kraich,gk,vergassola,dada,bbw,bif_magnetico}. \\
Another interesting question, opened by the failure of Kolmogorov
dimensional  prediction, consists in the possibility
that local observable with the same physical dimensions but with
 different tensorial structures have different scaling properties. 
About this point, there are some  experimental and numerical
evidences  regarding  different possible anomalous
 behaviour
of longitudinal and transversal structure functions \cite{sreene}
 even in isotropic turbulence. Somehow  related to this issue is also
 the apparent different anomalous scaling between  the coarse grained
 averages of dissipative quantities  like
enstrophy and energy dissipation \cite{chen1}. On the other hand,
 on the basis of a SO(3) decomposition  of velocity correlation functions,
other authors \cite{alp} claim that    the 
supposed  different scaling of quantities like transversal and longitudinal
 structure functions can only be due to spurious sub-leading
non-isotropic effects, i.e. in isotropic high-Reynolds numbers all components
 of the same tensorial observable should have the same -maybe anomalous-
 scaling behaviour. A first numerical support to this claim
has been presented in the analysis of a channel flow simulation
in \cite{prl1}. \\ Even more complex is the situation when multi-point 
pressure correlations is
involved \cite{iso1,hb97,press2}.  Dimensionally speaking pressure is 
just a velocity squared, and Kolmogorov-like argument can be  easily 
generalized 
to the case of pressure structure functions, $F_q(r)$. Indeed, a simple 
application of dimensional analysis leads to \cite{yaglom}:
\begin{equation}
F_q(r) \equiv \la\lab P(x)-P(x+r)\rab^q\ra \sim \epsilon^{2q} r^{2q}
\label{press}
\end{equation}
where as usual, all distances are supposed to belong to the inertial range 
of scales.  Of course, intermittency will also affect pressure scaling. By
 following the straightforward hypothesis that pressure can be treated as a
 velocity squared one would be tempted to 
assign  the same intermittency exponents of the velocity field to the 
pressure scaling, i.e. to replace   (\ref{press})  with 
\begin{equation}
F_q(r) \sim  r^{\zeta_v(2q)}
\label{press1}
\end{equation}
This prediction is just a simple consequences of the assumptions that all 
velocity correlations have the same scaling behaviour supposed that all 
distances involved are in the inertial range and  that the statistics is 
locally isotropic. Such a dimensional ansatz has been
questioned on the basis of a phenomenological argument in  \cite{iso1},
 some numerical 
support to the latter argument 
 have been recently presented in \cite{press2}. \\
In this paper we will mainly present some numerical evidences that indeed
the dimensional ansatz (\ref{press1}) is correct. It is well known that this
must be the case at least for 
 for $q=2$ in (\ref{press1}). In this case,  there exist
 an exact relation  \cite{hill} which connect the scaling of the second order
 pressure structure function with a linear integral combination of 
fourth-order velocity structure functions. The problem is  if the 
exact result can be simply extrapolated to other 
pressure-dependent observable and, in the case, 
how strong finite-Reynolds effects can be.
 Indeed, one may argue that
 pressure feels
strongly non-local effects, being just the inversion of the Poisson  problem 
$\Delta P = -\partial_i \partial_j v_i v_j$, and therefore the  assumptions of 
independence from large scales and/or from boundary conditions
 may  not be  satisfied even at very high Reynolds numbers. 
 Indeed, to our knowledge, 
neither experimental studies  nor numerical simulations have ever been able 
to make a firm quantitative statement about pressure scaling properties
 \cite{nc98,hb97,press2,lesieur}. In this paper we 
show that it is possible
to find a set of velocity-pressure observable which have indeed a quite good
scaling behaviour in agreement with the dimensional ansatz (\ref{press1}) 
also at moderate Reynolds numbers.\\
 Scaling  in turbulence is particularly
difficult to test in both experiments and numerical simulations. Experiments
 reaches high Reynolds numbers by  paying the price to have  
 a very limited set of information on the whole velocity fields,
 typically only a  long
time series of one velocity components in a few spatial points.
 Moreover, in most cases, there
is not a  precise control of the degree of isotropy and homogeneity in the 
flow. On the other hand, numerical simulations have a
 perfectly controlled  lay out, 
the velocity field is  exactly known at any point,
  but the maximum reachable Reynolds number is still 
order of magnitude smaller then in typical experiments \cite{chen1024}.
 \\ Nevertheless, numerical 
simulations, if exploited in a clever
way, are the only tool where complex measurements can be performed. Therefore, 
questions like the dependency of scaling properties from the tensorial 
nature of the observable
can, up to know, be investigated only in numerical data base. \\ 
In this paper, we present a detailed analysis of pressure scaling and 
pressure-velocity correlations scaling in a set of moderate  Reynolds number
 simulations. \\
Starting from the  analysis of the  energy transfer in real space 
 we propose a set of 
 pressure-velocity 
observable which show  better scaling properties then the usual 
pressure structure functions. We presents  {\it quantitative }
 evidences that indeed, while pressure structure functions are strongly 
affected from Reynolds numbers effects, the  pressure-velocity correlation
 functions we investigated  have a fairly good scaling behaviour, 
even at modest Reynolds numbers, in  agreement
with the hypothesis that pressure "behaves" like a velocity squared. In
 order to understand whether the bad scaling behaviour detected in the
pure-pressure structure functions is due to spurious anisotropic sub-leading 
effects we also present some 
results on the SO(3) decomposition of the pressure field. \\ The paper is 
organized as follows. In section 2 we summarize the known analitycal result
 which connect the second order pressure structure function to the integral 
linear combination of fourth-order velocity correlations and the experimental 
and numerical attempts to test the relation. In section 3 we introduce the 
set of pressure-velocity correlations which 
should have a better scaling properties on the basis of a simple argument
 based on the
energy transfer  of Navier-Stokes equations in the real space. 
In section 4
 we present the analysis of our numerical data base. In section 5
we briefly comment on the analysis of non isotropic fluctuations.  Conclusions 
follow in section 6.

\section{Pressure structure functions}
Under the assumptions of local  isotropy, local homogeneity, incompressibility,
and by use of Navier-Stokes equation, one can relate the second
order pressure structure functions, $F_2(r)$, to some fourth-order
velocity structure functions \cite{hill}. Namely:
\begin{eqnarray}
F_2(r) \equiv -\frac{1}{3} D_{1111}(r)  +  \frac{4}{3} r^2 \int_r^{\infty}
y^{-3}\left[D_{1111}(y)+D_{\beta\beta\beta\beta}(y)-6
D_{11\gamma\gamma}\right] dy \nonumber  \\
 +  \frac{4}{3}  \int_r^{\infty}
y^{-1}\left[D_{\beta\beta\beta\beta}(y)-3
D_{11\gamma\gamma}\right] dy 
\label{s4p2}
\end{eqnarray}
where the fourth-order structure function is 
$$D_{ijkl}(\vec{r})\equiv \la(u_i-u_i')(u_j-u_j')(u_k-u_k')(u_l-u_l')\ra$$
 and where for simplicity we have used primed variables to express velocities at the position $\vec{x'}=\vec{x}+\vec{r}$ and where $i,j,k,l$
is $1$ if the velocity component is parallel to the separation
vector, $\vec{r}$, and $2,3$ otherwise. Subscripts $\beta,\gamma$
denote either $2$ or $3$. Of course, (\ref{s4p2}) implies that 
whenever the fourth-order structure functions entering in the above
expressions are all dominated by the inertial-range intermittent 
scaling behaviour, $D_{i,j,k,l}(r) \sim r^{\zeta_v(4)}$, then
also the second order pressure structure functions should scale with the 
exponent $\zeta_p(2) = \zeta_v(4)$. Relation (\ref{s4p2}) have been carefully
tested in numerical simulations  without any appreciable deviations
\cite{hb97,nc98}. Nevertheless, the overall scaling behaviour 
of the pressure structure function is very poor. 
 Similarly, the analysis 
of experimental data \cite{nc98} does not show any power law behaviour
for the pressure structure functions even if the Reynolds number 
 was extremely high 
($Re_{\lambda} \sim 10000$). In the latter case, authors tried to explain
the difference  between pressure scaling quality and velocity 
scaling quality by invoking a possible different scaling
for the different velocity correlations entering in the RHS 
of (\ref{s4p2}), leading
to the final prediction that pressure structure functions is made in terms
of different power law contributions with slightly different exponents. 
The resulting superposition of power laws would be the responsible
of the poor observed scaling behaviour. This statement would anyhow
contradict the theoretical prediction made in terms of the SO(3) decomposition
which forbids different component of the same tensorial observable to
scale differently in a isotropic ensemble. \\
Another  interesting remark consists in the strong cancellation among
the different contribution of (\ref{s4p2}) observed in numerical simulations
\cite{hb97,nc98}: the LHS of (\ref{s4p2}) is more then an order of magnitude
smaller than the single different contributions entering in the RHS. 
One, cannot exclude apriori the possibility that there exist
an almost perfect cancellations of all leading scaling 
terms of all contribution appearing in the RHS of  (\ref{s4p2}), 
even if such a  perfect cancellation would call for some unknown physical
interpretation.\\ More probably, the cancellation is not perfect  but strong
enough to hide completely the pressure scaling at the available experimental
and numerical Reynolds numbers. \\ On the other hand,
the possibility that pressure-increments behave as
velocity-increments, $\delta P \sim \delta v$,
 instead than as a velocity-increment squared has been recently proposed
\cite{press2}.
 This would violate 
the exact results previously reported and therefore cannot be correct
unless strong anisotropic effects are present at all scales. \\
 In order to better assess the pressure
statistical properties we present in the following,
 some results for  pressure-dependent observable. This observable
do  not posses the strong cancellation properties showed by
structure function.  

\section{Pressure-velocity correlation}

Let us start by looking at the energy balance inside any volume $V$ 
of the flow. From the Navier-Stokes eqs we obviously have: 
\begin{equation}
\partial_t E_V + 
\int_V v_i(\bx)v_j(\bx)\partial_jv_i(\bx) d\bx  + 
\int_V v_i(\bx)\partial_i P(\bx) d\bx  = \nu \int_V v_i(\bx)\Delta v_i(\bx)
d\bx 
\label{en_ba}
\end{equation}
where with $E_V= 1/2 \int_V d \bx v_i v_i$
 we denote the total energy in the sub-volume $V$.  
%Let us now average over many realization without necessarly averaging
%over the whole space. We also assume stationarity and we neglect 
%the time derivative in what follows.
Let us notice that the two  terms in the LHS of (\ref{en_ba}) can be obviously written as the fluxes  across the boundaries of $V$ by using Gauss theorem:
\begin{equation} 
\int_V v_i(\bx)v_j(\bx)\partial_jv_i(\bx) d\bx \equiv \int_{\Omega_V}
d\Sigma n_j v_j(\bx) v^2(\bx) \equiv \Phi_{\Omega_V} ({\bf v} v^2)
\end{equation}
\begin{equation} 
\int_V  v_i(\bx)\partial_i P(\bx) d\bx \equiv \int_{\Omega_V}
d\Sigma n_i v_i(\bx)  P(\bx) \equiv \Phi_{\Omega_V}({\bf v}  P)
\end{equation}
where with $n_i$ we denote the versor perpendicular to the
 infinitesimal  surface on the boundaries
 of $V$. Rewritten in this way, relation (\ref{en_ba}) is just a
simple restatement of the conservation of energy: 
the total energy change inside a volume is given by the flux across
the volume surface and by the energy dissipation
 inside the volume. Let us now use this simple fact in order to 
extract some useful guess about scaling properties of velocity-pressure
correlations. Let us consider a very particular class of volume $V$, i.e.
a cylinder with an infinitesimal squared basis of surface $\epsilon$ and with 
a finite axis in the direction of $\bf{R}$. In the limit when the
basis becomes smaller and smaller, the flux across the lateral
sides goes to zero because contributions from two opposite  walls
are equal but with different sign. The only contributions to the 
total flux come from the two infinitesimal basis and can be 
written as
\begin{equation}
\Phi_{\Omega_V}({\bf v} v^2) = \epsilon( v_i(\bx)v^2(\bx)-
v_i(\bx +\bR)v^2(\bx+\bR))\hat{R}_i
\end{equation}
for the flux involving the velocity correlation and as
\begin{equation}
\Phi_{\Omega_V}({\bf v} P ) = \epsilon( v_i(\bx)P(\bx)-
v_i(\bx +\bR)P(\bx+\bR))R_i
\end{equation}
for the flux involving pressure-velocity correlations. 
In both cases we have exploited 
the fact that the two infinitesimal basis are
centered in $\bx$ and in $\bx +\bR$, i.e. their versor is
oriented along  
 $\hat{R}$. Similarly the two volume integral giving the time
variation of the total energy and the energy dissipation becomes two
linear integral times the infinitesimal basis area $\epsilon$:
$\partial_t (\epsilon \int_{\bR} ds v^2(\bx))$ and 
$(\epsilon \int_{\bR} ds  v_i(\bx)\Delta v_i(\bx))$ respectively,
 where with $ds$ we parametrized the segment
going from $\bx$ to $\bx+\bR$. \\Let us now 
assume that all the four 
observable entering in the energy balance have the same statistical
behaviour.
 This is somehow a ``local Kolmogorov refined hypothesis'': we link the
scaling of the local energy dissipation to the scaling of
some particular third order velocity correlation and to the
scaling of a velocity-pressure correlation. The claim is therefore
that the particular structure functions emerging from our flux 
analysis should scale exactly like the coarse grained energy dissipation,
i.e. should have an anomalous scaling  like the usual
longitudinal structure functions which  satisfy
the original Kolmogorov Refined Hypothesis. In the next section
we present some numerical data in support of this claim.

\section{Numerical Analysis} 
The  data set we are going to analyze has been obtained from
a direct numerical integration of NS eqs using a pseudo-spectral method
with dealiasing on a grid of $128^3$  points. The forcing was
implemented isotropically on all wavevectors with $|k|<1$ such as to 
enforce the $k^{-5/3}$ spectrum at small wavectors \cite{forcing}.
We have analyzed about 
100 configurations stored each eddy-turn over time. The
simulation has a $Re_{\lambda}=70$.\\
Let us denote with:
\begin{eqnarray}
S_{q}^{vv^2}(R)& = &\la\lab\la v_i(\bx)v^2(\bx)-
v_i(\bx +\bR)v^2(\bx+\bR)\ra\hat{R}_i\rab^{q/3}\ra \nonumber \\
S_{q}^{vP}(R)& = &\la\lab\lp v_i(\bx)P(\bx)-
v_i(\bx +\bR)P(\bx +\bR)\rp\hat{R}_i\rab^{q/3}\ra 
\label{sf3}
\end{eqnarray}
the two different structure functions which can be made in terms of the 
two flux quantities defined in the previous section.
 Let us notice that in (\ref{sf3}) both 
$ S_{q}^{vv^2}(R)$ and $S_{q}^{vP}(R) $ have been defined as the $q/3$ power
of the original  fluxes such as to have the same dimensions of $S^v_{q}(R)$.\\
Let us start by showing in Fig.1 
 the strong cancellation effects present in the pressure
structure functions $F_{q}(R)$ with respect 
to the velocity longitudinal structure functions with the 
same physical dimensions $S_{2q}^v(R)$. In Fig. 1 we show 
the log-log plot of $F_1(R)$ and of $S_2(R)$,
 as one can see the overall amplitude
of pressure fluctuations is about 
 one order of magnitude  smaller  than the velocity fluctuations. This 
is just to confirm that pressure by itself is a much weaker signal than
the usual velocity correlations.\\ As one can see in Fig. 1 
the scaling is quite poor, as one can expect in any DNS. As usual, in order
to extract quantitative statement about scaling exponents one has
to exploit the Extended Self Similarity (ESS) property enjoyed
by homogeneous and isotropic turbulent flows \cite{ess}. ESS consists
in looking for relative scaling of two different observable. Usually,
one takes two structure functions of two different orders, i.e. 
in the case of longitudinal structure functions 
$S_{q}^v(R) \sim \left[S_{q'}^v(R)\right]^{\zeta_v(q)/\zeta_v(q')}$. \\
Let us now define the same relative scaling for the two generalized
structure functions defined in (\ref{sf3}).
\begin{equation}
S_{q}^{vv^2}(R) \sim \left[S_{q'}^{vv^2}(R)\right]^{\frac{\zeta_{vv^2}(q)}
{\zeta_{vv^2}(q')}}, \;\; S_{q}^{vP}(R) 
\sim \left[S_{q'}^{vP}(R)\right]^{\frac{\zeta_{vP}(q)}{\zeta_{vP}(q')}}
\label{ess_gen}
\end{equation}
In Figs. 2,3,4 we show the ESS plot for the two generalized
structure functions and for the usual longitudinal structure functions 
respectively, with  $q=1,\,q'=2$.  
 As one can see the all  ESS plots showed a
scaling behaviour consistent with the usual homogeneous and isotropic 
high Reynolds value  which give for the relative  exponents: 
$\zeta_v(2)/\zeta_v(1) = 1.92 \pm 0.02$ \cite{ess}. %Previous \cite{exponents}
 Similar agreements are found 
for higher order moments (not showed). \\ The scaling
ansatz assumed for the generalized structure functions seems therefore
quite well satisfied. These findings support the fact  that pressure
does not behave abnormally as far as its ``dimensional'' scaling 
properties are concerned. Indeed, pressure-velocity correlations
behaves exactly like velocity-velocity correlations once pressure
is counted as a ``velocity squared''. Nevertheless,
The only realistic way to perform  a quantitative statement about 
scaling exponents is to study the logarithmic local slopes of
(\ref{ess_gen}). Only when  logarithmic local slopes show a fairly
constant behaviour one can really speak about scaling. In Fig. 5
we show the logarithmic local slopes of (\ref{ess_gen}) for $q=1,\,q'=2$ 
together with the corresponding quantities measure for the 
longitudinal velocity structure functions. In order to show that 
the hypothesis that pressure-increments behave as a  linear
velocity-increment, $\delta P \sim \delta v$,   is definitely
ruled out by our data we also show in Fig. 5 the logarithmic local slope
of the ESS applied to pressure structure-functions for $q=1,q'=2$.
As one can see, while the three slopes measured on the flux structure
functions and on the longitudinal structure functions agree perfectly
with the high-Reynolds numbers measurements, the pure-pressure
structure functions is definitely much poorer. In Fig. 6 we plot
the same of Fig. 5 but for a different choice of moments, $q=2,\;q'=4$,
for both fluxes and longitudinal structure functions and with 
$q=1,q'=2$ for the pure pressure structure functions.  In this way
we are comparing  quantities with exactly the same dimensional properties. 
Again, while the flux-made structure functions, $S^{vP}_q(R)$, $S^{vv^2}_q(R)$
 and the longitudinal structure
functions, $S^v_q(R)$ have the same local slope the pure-pressure 
result obtained on the ESS of $F_q(R)$ shows a poorer and
different scaling. \\ A few comments are now in order. On one hand,  we see
from Figs. 5 and 6 that the simple local-refined Kolmogorov hypothesis
derived in the previous section is correct, i.e.  fluxes (\ref{sf3}) have
the same scaling properties of the usual longitudinal structure
function in homogeneous and isotropic turbulence, confirming
that these observable with the same physical dimensions and built
in terms of local field increments scale in the same way.
 On the other hand, (see Fig. 6)
pure-pressure structure functions seem to violate the previous statement  
despite of the fact that in this case there even exist  an
exact result (\ref{s4p2}) supporting  it.
 Why pure-pressure correlations  show this strong
deviation from the straightforward dimensional estimate? One possible 
explanation is connected to the -possible- lack of isotropy in the statistics.
Any isotropically forced DNS is affected by possible
 non-isotropic fluctuations induced by the 
discretization of the numerical grid. In the following section 
we have analyzed non-isotropic effects on both velocity and pressure
fluctuations. 
\section{Anisotropic effects}
The exact relation which connect the second order 
pressure structure function with a linear integral combination
of fourth order velocity structure function (\ref{s4p2}) is correct
only in the isotropic and homogeneous case. In order to test the
degree of isotropy of our simulation we have proceeded in a systematic
decomposition in terms of the irreducible representations of the SO(3)
symmetry group \cite{alp,prl1,itamar1}. The SO(3)
decomposition is particularly simple to apply to scalar observable,
i.e. observable whit all vectorial indexes contracted, like 
pressure structure functions or longitudinal structure functions 
on the kind analyzed in this work. In these case, the SO(3) decomposition
is nothing but a decomposition in spherical harmonics. For example, the 
longitudinal structure functions, $S_q^v(\bR) 
=< (v_i(\bx+\bR)-v_i(\bx))\hat{R}_i)^q> $,
 can be decomposed as:
\begin{equation}
S_q^v(\bR) \equiv \sum_{jm} S_q^{jm}(R) Y_{jm}(\hat{\bR}) 
\label{so3}
\end{equation}
where now, we have explicitly considered the possibility that the undecomposed
structure functions depend on the whole vector $\bR$ and not only
on its magnitude as in the previous sections when  isotropy was assumed. 
The coefficient of the decomposition, $S_q^{jm}(R)$ depend only on the
magnitude of $R$ and on the two ``quantum'' numbers $j,m$ which labels
the properties under rotations of the $Y_{jm}$ eigenfunction. 
Obviously, in the case of perfect isotropy we would have only one
projection alive, i.e. the projection on $Y_{00}$. The SO(3) decomposition
here summarized as been already used in some experimental and numerical
data analysis to properly disentangle the anisotropic effects
 from the isotropic
ones \cite{alp,prl1,itamar1,bif_magnetico}.  In our case, the relative
amplitudes of $S_q^{00}(|R|)$ with respect to the anisotropic fluctuations
$S_q^{jm}(|R|)$ with $j>0, -j \leq m \leq j$, gives a direct 
quantitative estimate of the degree of anisotropic fluctuations
for any scale $|R|$.  \\
In Fig. 7 we show the log-log plot of the undecomposed second order 
longitudinal structure functions and of its projection on the fully-isotropic
eigenfunction $Y_{00}$. As it is possible to see, despite of the
isotropic forcing used in the simulation, the finite-size effects introduced
by our computational grid are quite important at large scales: the 
projection, $S_2^{00}(|R|)$, shows a definitely better scaling than the
undecomposed structure function already as a functions of the
real separation $R$, i.e. without using ESS. This dramatic effect was already
observed in a similar application to the decomposition of velocity
fluctuations inside a channel \cite{prl1}. Fig. 7 definitely show that
the SO(3) decomposition can help in cleaning scaling properties 
also in ``quasi-isotropic'' simulations. \\
On the other hand, the situation is quite different when the same decomposition
is applied to the pressure structure functions:
\begin{equation}
F_q^v(\bR) \equiv \sum_{jm} F_q^{jm}(R) Y_{jm}(\hat{\bR}) 
\label{Pso3}
\end{equation}
Let us notice that pressure is a quasi-isotropic  observable  
also for strong anisotropic velocity configuration. Indeed, being the 
solutions of a Poisson problem, pressure is always an average of
velocity fluctuations on all spatial directions. This simple 
considerations is perfectly verified on our numerical simulation. In Fig. 8
we show the undecomposed second-order pressure structure function
together with its projection on the fully isotropic harmonics, $F_q^{00}(R)$.
One can hardly detect any differences, suggesting tha anisotropic
fluctuations cannot be the responsible for the poor
scaling observed in the previous sections for the pure-pressure
structure functions. \\ The only possibility to reconcile the 
exact result (\ref{s4p2}) with the poor scaling agreement
between pressure structure functions and velocity structure functions
is, in our opinion,  to invoke strong Reynolds effects. Indeed, the resolution
of the Poisson problem certainly introduces strong non-local
effects on the statistics. Non-locality may also translate in strong 
long-range effects in the Fourier space as far as the importance
of boundary conditions and forcing on the inertial range properties are
concerned. If this is correct, there are not reason to expect good
scaling properties for pure-pressure correlations, unless Reynolds number
is high enough to recover also in laboratory experiments 
an almost-``infinite'' inertial range extension.
\section{Conclusions}
We have analyzed some  pressure and pressure-velocity correlations
in a Direct Numerical Simulations at moderate Reynolds numbers. 
We have derived on the basis of a simple analysis of energy transfer
properties across any sub-volume in the real space what we call
a ``local''-Refined Kolmogorov Hypothesis. We have identified a set
of pressure-velocity correlations which should have a good scaling 
behaviour because connected via the local-RKH to the scaling of the
energy dissipation coarse grained on inertial range scales. \\We have 
showed that our scaling hypothesis is well verified, while pure pressure
correlations feels strong Reynolds effects. According to our analysis
pressure scaling is perfectly determined by the dimensional
assumption that pressure behaves as a ``velocity squared'', unless the
finite-Reynolds effects are overwhelming. We do not find any sign
which could support the fact that pressure differences 
behave as velocity differences as proposed in \cite{press2}.\\
We have also applied the SO(3) decompositions to the pressure
structure functions in order to show that poor scaling properties
showed by pure-pressure structure functions are not connected
to anisotropic fluctuations. \\ 
We acknowledge some help by I. Mazzitelli in the SO(3) analysis. LB
and FT have  been partially supported by INFM (PRA-TURBO) and by the EU 
contract FMRX CT98-0175,

\newpage
%%%%%%%%%%%%%%%%%%%%%%%%
\begin{figure}
%\narrowtext
\epsfxsize=16truecm
\epsfysize=10truecm
\epsfbox{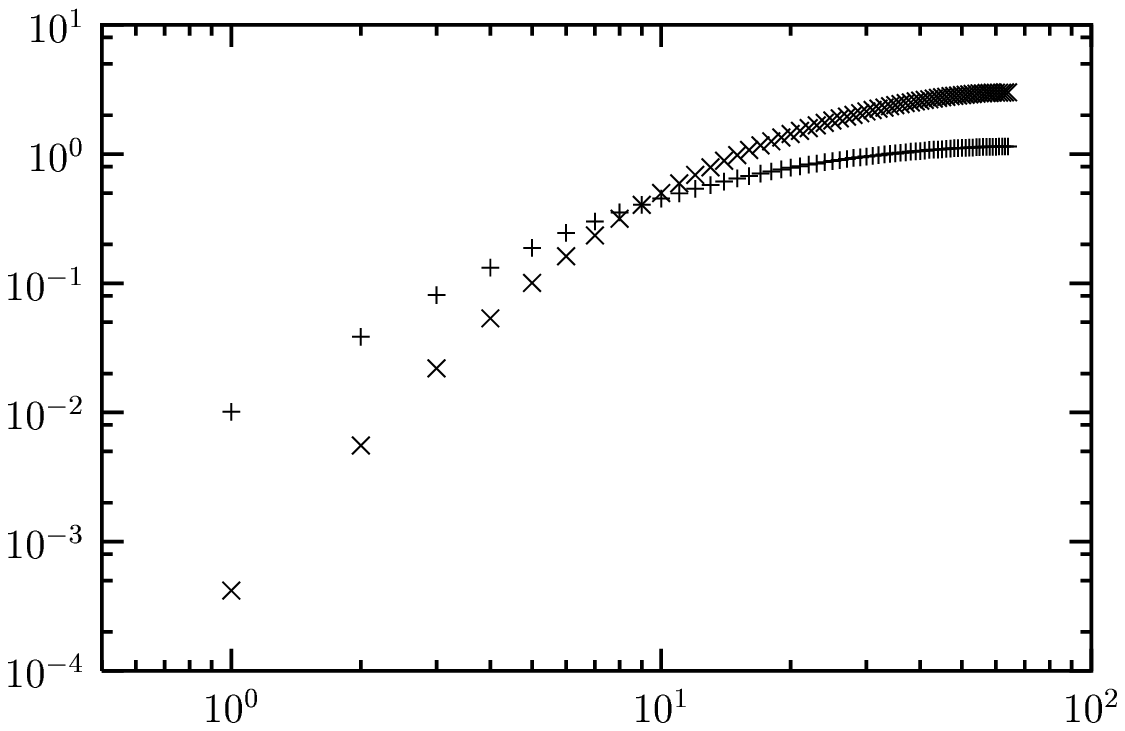}
\caption{Log-log plot of $F_1(R)$ $+$, and of $S_2(R)$, $\times$. Notice
that the pressure structure function is about one order of magnitude 
smaller than the velocity structure function at large scale. }
\end{figure}
%%%%%%%%%%%%%%%%%%%%%%%%
%%%%%%%%%%%%%%%%%%%%%%%%
\begin{figure}
%\narrowtext
\epsfxsize=16truecm
\epsfysize=10truecm
\epsfbox{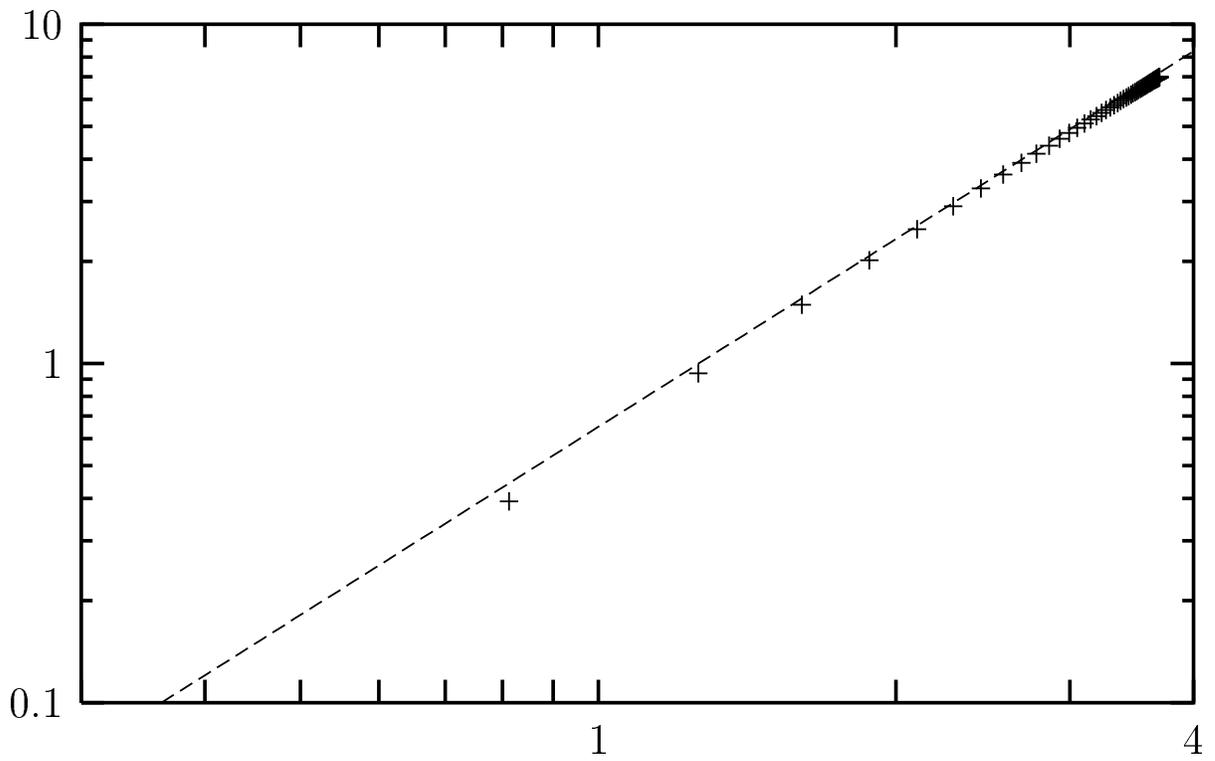}
\caption{ESS log-log plot of $S^{vv^2}_2(R)$ vs $S^{vv^2}_1(R)$,
 superimposed 
is the straight line with the isotropic and homogeneous high-Reynolds slope
$\zeta_2^{v}/\zeta_1^v = 1.92$.}
\end{figure}
%%%%%%%%%%%%%%%%%%%%%%%%
%%%%%%%%%%%%%%%%%%%%%%%%
\begin{figure}
%\narrowtext
\epsfxsize=16truecm
\epsfysize=10truecm
\epsfbox{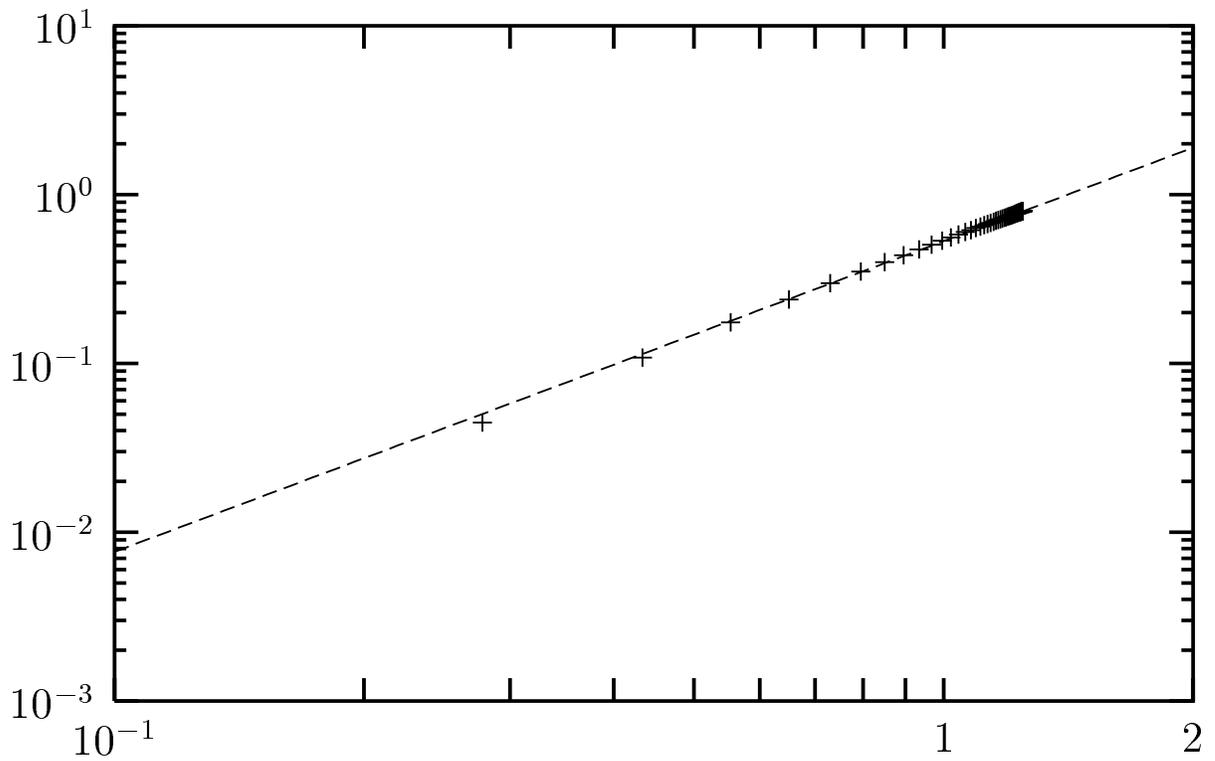}
\caption{The same as in Fig. 2 but for  $S^{vP}_2(R)$ vs $S^{vP}_1(R)$.}
\end{figure}
%%%%%%%%%%%%%%%%%%%%%%%%
%%%%%%%%%%%%%%%%%%%%%%%%
\begin{figure}
%\narrowtext
\epsfxsize=16truecm
\epsfysize=10truecm
\epsfbox{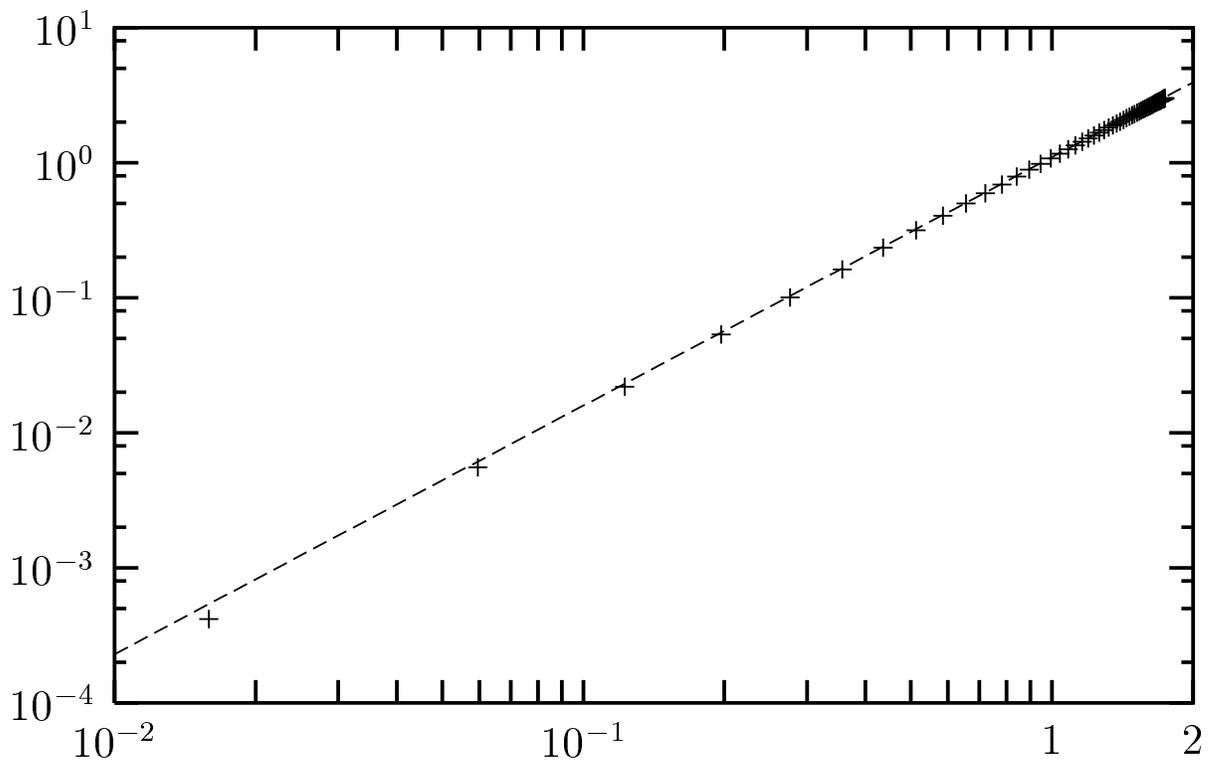}
\caption{The same as in Fig. 2 but for the longitudinal
structure function, $S^v_2(R)$ vs $S^v_1(R)$.}
\end{figure}
%%%%%%%%%%%%%%%%%%%%%%%%

%%%%%%%%%%%%%%%%%%%%%%%%
\begin{figure}
%\narrowtext
\epsfxsize=16truecm
\epsfysize=10truecm
\epsfbox{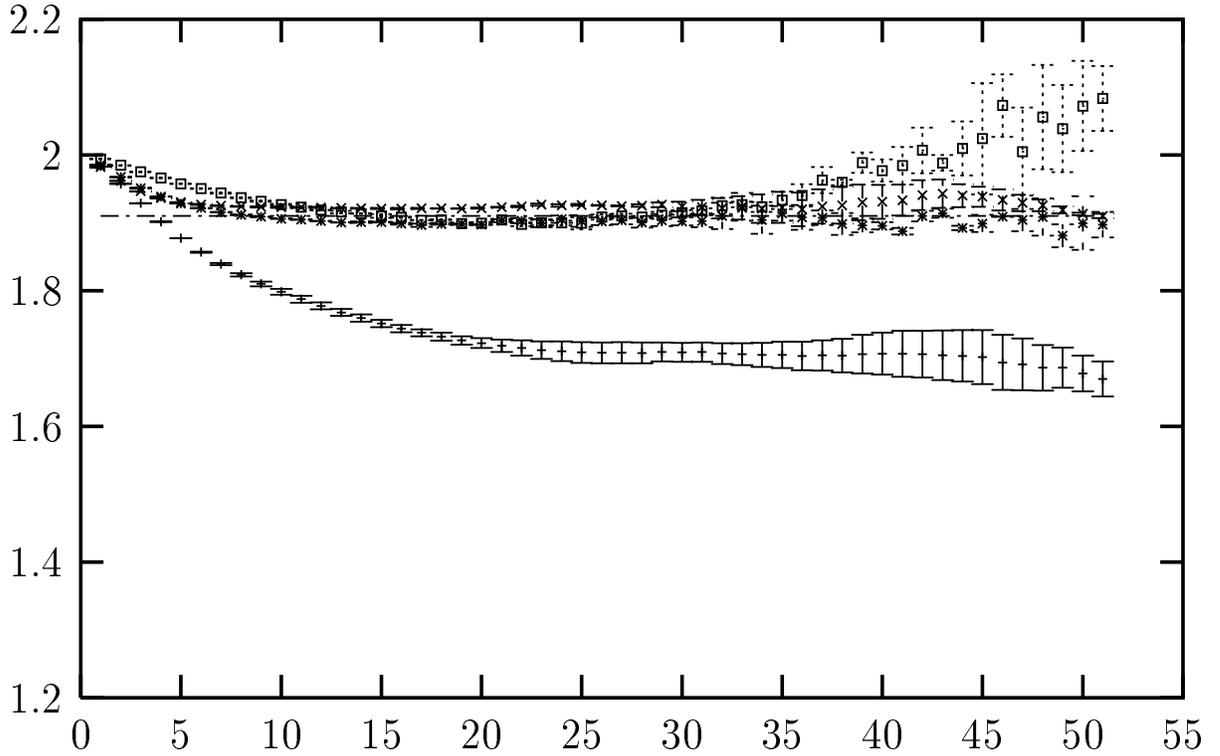}
\caption{Logarithmic local slope of: ($\ast$) 
 $S^{vv^2}_q(R)$ vs $S^{vv^2}_{q'}(R)$; ($\Box$)
$S^{vP}_q(R)$ vs $S^{vP}_{q'}(R)$; ($\times$)  $S^{v}_q(R)$ vs $S^{v}_{q'}(R)$,
for $q=2$ and $q'=1$. 
Notice that all the above values are in perfect agreement with 
the high-Reynolds number value $1.92$ (straight line), while the
logarithmic local slope for  the pure-pressure structure
functions  $F_2(R)$ vs $F_1(R)$, ($+$) is different.
The error bars are estimated by looking at the fluctuations over the first
half and the second half of the whole set of configurations.}  
\end{figure}
%%%%%%%%%%%%%%%%%%%%%%%%
%%%%%%%%%%%%%%%%%%%%%%%%
\begin{figure}
\epsfxsize=16truecm
\epsfysize=10truecm
\epsfbox{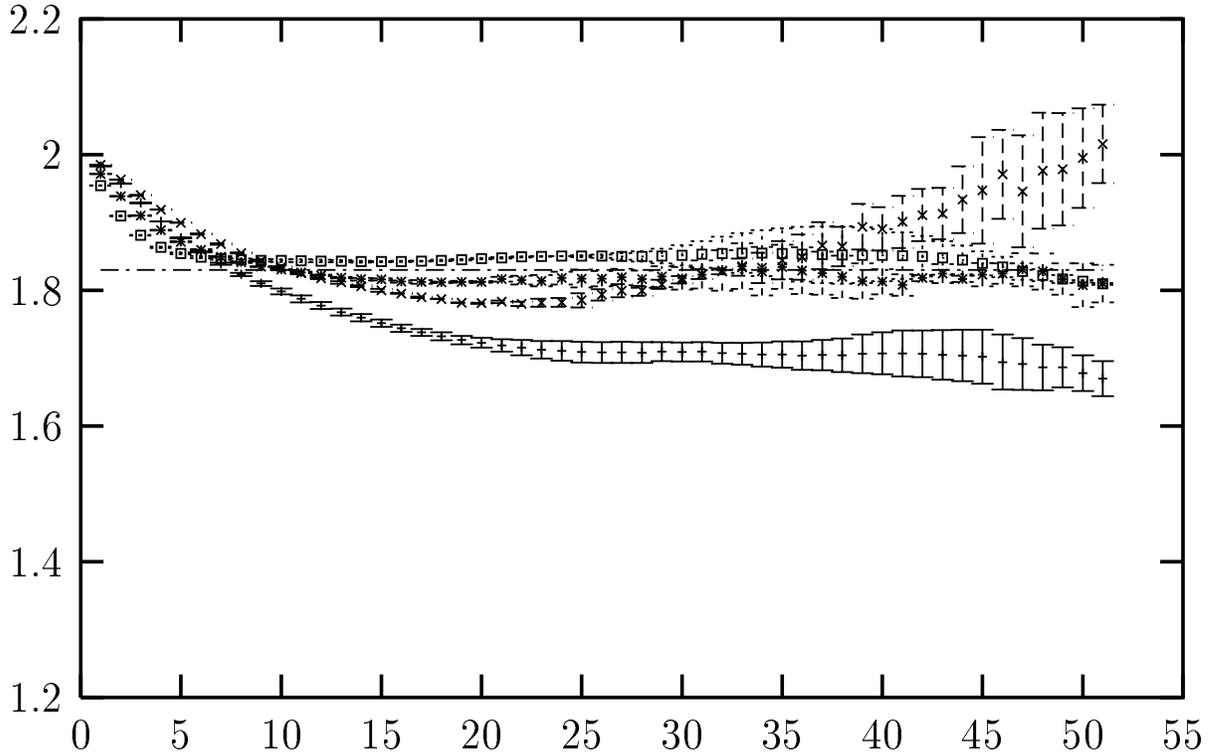}
\caption{Logarithmic local slopes of:  ($\Box$) 
$S^{vv^2}_q(R)$ vs $S^{vv^2}_{q'}(R)$; ($\ast$)
$S^{vP}_q(R)$ vs $S^{vP}_{q'}(R)$; ($\times$)  $S^{v}_q(R)$ vs $S^{v}_{q'}(R)$,
for $q=4$ and $q'=2$.
Notice that, as in Fig. 5, the flux-based structure
 functions have the same scaling behaviour
of the longitudinal structure function in agreement with the 
high-Reynolds regime (straight line). Here we present a comparison with
the Pressure structure function with the same physical dimensions
of the flux-based structure functions, i.e. $F_2(R)$ vs $F_1(r)$, ($+$),
still  the pure-pressure 
structure function seems to have a different local slope.}
\end{figure}
%%%%%%%%%%%%%%%%%%%%%%%%

%%%%%%%%%%%%%%%%%%%%%%%%
\begin{figure}
\epsfxsize=16truecm
\epsfysize=10truecm
\epsfbox{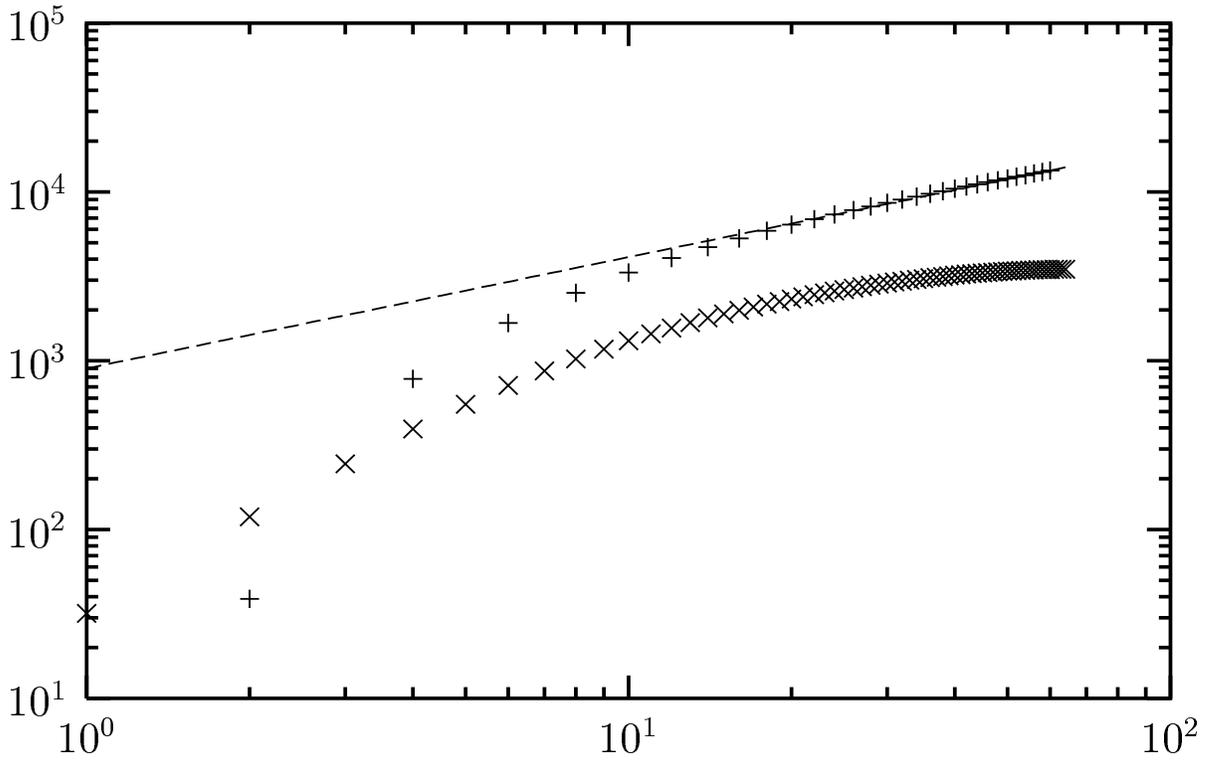}
\caption{Undecomposed second order velocity structure functions
$S_2^v(\bR)$ measured on the plane $x-y$, ($\times$);
 and the  projection, $S_2^{00}(R)$ ($+$),
 on the isotropic eigenfunction. The straight
line has the high-Reynolds slope $\zeta_2^v=0.7$. Notice that  already 
in the $R$-space, the SO(3)
 decomposition improve the overall scaling behaviour. The two curves have been 
shifted along the $y$ axis  for the sake of presentation.}
\end{figure}
%%%%%%%%%%%%%%%%%%%%%%%%

%%%%%%%%%%%%%%%%%%%%%%%%
\begin{figure}
%\narrowtext
\epsfxsize=16truecm
\epsfysize=10truecm
\epsfbox{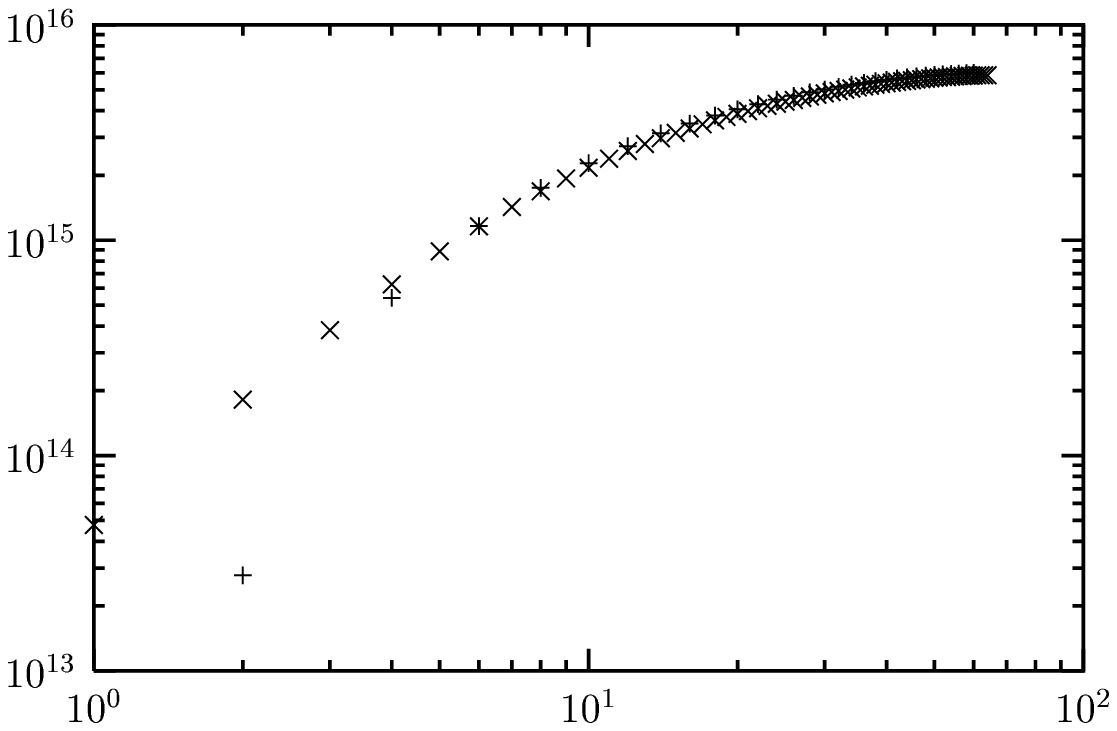}
\caption{Undecomposed second order pressure structure functions
$F_2({\bf R})$ measured on the plane $x-y$, ($\times$);
 and the projection, $F_2^{00}(R)$ ($+$),
 on the isotropic eigenfunction.
 Notice that, at difference from Fig. 7, here  
 the decomposed and undecomposed  structure functions are almost identical,
indicating a high degree of isotropization in the pressure statistics.}
\end{figure}
%%%%%%%%%%%%%%%%%%%%%%%%


\begin{thebibliography}{99}

\bibitem{frisch}  U. Frisch, ``Turbulence: the Legacy of A.N. Kolmogorov
(Cambridge University Press, Cambridge, UK, 1995).

\bibitem{iso1}M. Nelkin, ``Universality and scaling in fully developed 
turbulence'', Advances in Physics, {\bf 43}, 143 (1994). 

\bibitem{iso2}K.R. Sreenivasan and R.A. Antonia,
``The phenomenology of small-scale turbulence''
 Annu. Rev. Fluid Mech. {\bf 29}
435 (1997).

\bibitem{prl1} I. Arad, L. Biferale, I.  Mazzitelli and I. Procaccia,
`` Disentangling scaling properties in anisotropic and
 inhomogeneous Turbulence''
Phys. Rev. Lett. {\bf 82}  5040 (1999). 

\bibitem{prl2} F. Toschi, G. Amati, S. Succi, R. Piva, R. Benzi,\\
``Intermittency and structure functions in channel flow turbulence'',\\
Phys. Rev. Lett. {\bf 82} (1999) 5044.

\bibitem{itamar1}I. Arad, B. Dhruva, S. Kurien, V.S.L'vov, I. Procaccia
and K.R. Sreenivasan, ``Extraction of anisotropic contributions in 
turbulent flows'' Phys. Rev. Lett. {\bf 81} 5330 (1998).

\bibitem{kraich} R.H. Kraichnan, ``Anomalous scaling of a 
randomly advected passive scalar'' Phys. Rev. Lett. {\bf 72} 1016 (1994)

\bibitem{gk} K. Gawedski and A. Kupiainen, ``Anomalous scaling of the passive scalar'' Phys. Rev. Lett. {\bf 75} 3834-3837 (1995).

\bibitem{vergassola} M. Vergassola ``Anomalous scaling for passively advected magnetic fields'', Phys. Rev. E {\bf 53} R3021-R3024 (1996).

\bibitem{dada} A. Lanotte and A. Mazzino ``Anisotropic nonperturbative zero modes for passively advected magnetic fields'' Phys. Rev. E {\bf 60} R3483-R3486 (1999).


\bibitem{bbw} R. Benzi, L. Biferale and A. Wirth 
``Analytic calculation of anomalous scaling in random
 shell models of passive
scalars''
Phys. Rev. Lett. {\bf 78}  4926 (1997).

\bibitem{bif_magnetico} I. Arad, L. Biferale and I. Procaccia, 
``Nonperturbative Spectrum of Anomalous Scaling Exponents 
in the Anisotropic Sectors of Passively Advected Magnetic Fields ``
Phis. Rev. E (1999) submitted.  

\bibitem{sreene} B. Dhruva, Y. Tsuji and K.R. Sreenivasan
``Transverse structure functions in high-Reynolds-number turbulence''
Phys. Rev. E {\bf 56} R4928 (1997).

\bibitem{chen1} S. Chen, K.R. Sreenivasan, M. Nelkin, and N. Cao ``Refined
similarity hypothesis for transverse structure functions in fluid turbulence''
Phys. Rev. Lett. {\bf 79} 2253 (1997). 

\bibitem{alp} I. Arad, V.S.  L'vov and I. Procaccia, ``Correlation functions in isotropic and anisotropic turbulence: the role of the symmetry group'' Phys. Rev. E {\bf 59} 6753 (1999).

\bibitem{nc98} M. Nelkin and S. Chen, ``The scaling of pressure in isotropic 
turbulence'', Phys. Fluids {\bf 10} 2119 1998. 

\bibitem{hb97} R.G. Hill and O. N. Boratav ``Pressure statistics for
locally isotropic turbulence'' Phys. Rev. E {\bf 56} R2363 1997. 

\bibitem{press2} N. Cao, S. Chen and G.D. Doolen ``Statistics and structures
of pressure in isotropic turbulence'' Phys. Fluids {\bf 11} 2235 (1999). 

\bibitem{yaglom} A.S. Monim and A.M. Yaglom ``Statistical Fluid Mechanics''
Vol. 2 M.I.T Press Cambridge (1975). 

\bibitem{hill}R. J. Hill and J. M. Wilczak ``Pressure structure functions
and spectra for locally isotropic turbulence'' J. Fluid Mech. {\bf 296}
247-269, 1995. 

\bibitem{vedula} P. Vedula and P.K. Yeung ``Similarity scaling 
of accelaration and pressure statistics in numerical 
simulations of isotropic turbulence'' Phys. Fluids {\bf 11} 1208 (1999). 

\bibitem{duady}S. Douday, Y. Couder and M.E. Brachet ``Direct 
observation of the intermittency of 
intense vorticity filaments in turbulence'' 
Phys. Rev. Lett. {\bf 67} 983 (1991).

\bibitem{forcing} M.R. Overholt and B.Pope,
    ``A deterministic forcing scheme for direct numerical 
simulation of turbulence'',  Computers \& Fluids {\bf 27} 1 1998

\bibitem{lesieur} M. Lesieur ``Turbulence in Fluids'',(third revised 
edition,  Kluwer Academic Publisher 1997).

\bibitem{chen1024} S.Y. Chen and X. Shan ``High resolution 
turbulent simulations using the connectin machine-2'' Compt. Phys. {\bf 6} 643
(1992).

\bibitem{ess} R. Benzi, S. Ciliberto, R. Tripiccione, C. Baudet, F. Massaioli
and S. Succi ``Extended Self Similarity in turbulent flows'' Phys. Rev. E
{\bf 48} R29 (1993).

%\bibitem{exponents} a ref where one can finds exponents measured at high
%reynolds numebrs, maybe one of the ref of roberto on ESS

\end{thebibliography}
\end{document}